\documentclass{article}

\usepackage{PRIMEarxiv}
\usepackage{multirow}
\usepackage{makecell}
\usepackage{amsmath}
\usepackage{afterpage}
\usepackage{rotating}
\usepackage{subcaption}
\usepackage{algorithm}
\usepackage{algpseudocode}
\usepackage{amsmath}
\usepackage{graphicx}

\usepackage[utf8]{inputenc} 
\usepackage[T1]{fontenc}    
\usepackage{hyperref}       
\usepackage{url}            
\usepackage{booktabs}       
\usepackage{amsfonts}       
\usepackage{nicefrac}       
\usepackage{microtype}      
\usepackage{lipsum}
\usepackage{graphicx}
\graphicspath{{media/}}     
\usepackage{float}
  
\title{Enhancing Knowledge Retrieval with In-Context Learning and Semantic Search through Generative AI}

\author{
  Mohammed-Khalil Ghali, Abdelrahman Farrag, Daehan Won, Yu Jin \\
  School of Systems Science and Industrial Engineering \\
  State University of New York at Binghamton \\
  Binghamton, NY, USA\\
  \texttt{\{mghali1, afarrag1, dhwon, yjin\}@binghamton.edu} \\
}

\begin{document}
\maketitle

\begin{abstract}
Retrieving and extracting knowledge from extensive research documents and large databases presents significant challenges for researchers, students, and professionals in today's information-rich era. Existing retrieval systems, which rely on general-purpose Large Language Models (LLMs), often fail to provide accurate responses to domain-specific inquiries. Additionally, the high cost of pretraining or fine-tuning LLMs for specific domains limits their widespread adoption.
To address these limitations, we propose a novel methodology that combines the generative capabilities of LLMs with the fast and accurate retrieval capabilities of vector databases. This advanced retrieval system can efficiently handle both tabular and non-tabular data, understand natural language user queries, and retrieve relevant information without fine-tuning. The developed model, Generative Text Retrieval (GTR), is adaptable to both unstructured and structured data with minor refinement. GTR was evaluated on both manually annotated and public datasets, achieving over 90\% accuracy and delivering truthful outputs in 87\% of cases. Our model achieved state-of-the-art performance with a Rouge-L F1 score of 0.98 on the MSMARCO dataset.
The refined model, Generative Tabular Text Retrieval (GTR-T), demonstrated its efficiency in large database querying, achieving an Execution Accuracy (EX) of 0.82 and an Exact-Set-Match (EM) accuracy of 0.60 on the Spider dataset, using an open-source LLM. These efforts leverage Generative AI and In-Context Learning to enhance human-text interaction and make advanced AI capabilities more accessible. By integrating robust retrieval systems with powerful LLMs, our approach aims to democratize access to sophisticated AI tools, improving the efficiency, accuracy, and scalability of AI-driven information retrieval and database querying.
\end{abstract}


\section{Introduction}
The ever-growing volume and diversity of academic literature and large databases present a formidable challenge for researchers and professionals seeking relevant insights and precise answers to specific questions. Traditional retrieval systems often fall short in addressing this complexity, leading to inefficiencies and suboptimal knowledge discovery. While Large Language Models (LLMs) have demonstrated remarkable capabilities in natural language understanding, their applicability in handling domain-specific knowledge and extensive context windows remains limited. This limitation hinders the effectiveness of existing systems in managing and retrieving information from large documents and tabular data.
In this context, the successful implementation of advanced retrieval systems can revolutionize the way researchers and professionals interact with large datasets, streamline information retrieval processes, foster knowledge discovery, and enhance the extraction of insights. This research endeavor aims to bridge the gap between the general-purpose capabilities of LLMs and the specific demands of domain-specific knowledge handling.
Two primary challenges underscore the gap in the current literature regarding domain-specific information retrieval. Firstly, existing retrieval systems predominantly rely on general-purpose LLMs, which lack the nuanced understanding of domain-specific contexts. As a result, these systems often provide inaccurate or inadequate responses to queries concerning documents and tabular data. The absence of specialized knowledge representations significantly hampers the ability of these models to navigate complex academic language and databases, thereby limiting the precision of the information retrieved.
Secondly, the high cost and resource-intensive nature of fine-tuning LLMs present substantial financial and computational barriers. Comprehensive model fine-tuning is often unfeasible for many research applications, rendering the deployment of highly specialized models impractical.
To address these challenges, there is a pressing need for an advanced retrieval system capable of efficiently grasping domain-specific knowledge within vast repositories of both structured and unstructured text. The primary objective of this research is to design an intelligent system that comprehends the context of user queries and retrieves relevant information with high precision and truthfulness. This system should circumvent the exorbitant costs associated with extensive model fine-tuning and pre-training.
This research paper proposes a novel approach that combines the capabilities of LLMs with vector databases to create a robust retrieval system. The approach involves developing methods to integrate domain-specific knowledge into the retrieval system without the need for extensive fine-tuning, which could involve leveraging pre-trained models and augmenting them with domain-specific embeddings or knowledge bases. Additionally, the system will utilize vector databases to efficiently index and retrieve information from large datasets. Vector databases enable the representation of documents and queries as high-dimensional vectors, facilitating rapid and accurate similarity searches.

\section{Literature review}
In contrast to prevailing assumptions, \cite{gu2021domain} highlights significant benefits of starting domain-specific language models from scratch, particularly in biomedicine. They establish state-of-the-art results across tasks, showing the efficacy of domain-specific pretraining. \cite{zheng2022pretrained} focuses on domain-specific pre-trained models in architecture, engineering, and construction (AEC) NLP tasks, revealing their potential for enhancing AEC industry information processing.
Language models (LLMs) have limitations in encapsulating application-specific information, especially for time-sensitive tasks like news question answering, due to evolving real-world contexts. Restricted proprietary datasets for LLM training due to privacy concerns further constrain their knowledge. While enhancing LLMs with external knowledge is popular, prior methods often require costly parameter fine-tuning as model sizes grow \cite{zhong2022training}. BERT, introduced by \cite{devlin2018bert}, employs bidirectional representations across layers, allowing fine-tuning for various tasks. BERT's success led to similar models \cite{liu2019roberta,yang2020textbrewer}. In-context learning aids tasks, though struggles with complex reasoning \cite{wei2022chain}. Recent explorations into prompting techniques, exemplified by \cite{wei2022chain}, and \cite{zhou2022least}, have unveiled that explicating reasoning steps in examples can amplify logical abilities of language models (LLMs). This approach, termed chain-of-thought (CoT) prompting, exhibits promise. However, its implementation necessitates manual input, entailing informative question selection and CoT annotation.
Pre-trained neural language models exhibit the remarkable capacity to acquire knowledge without external memory, utilizing parameterized implicit knowledge bases \cite{roberts2020much}. Despite progress, limitations persist, encompassing memory expansion, transparent explanations, and "hallucinations" \cite{marcus2020next}. Some hybrid models \cite{guu2020retrieval,ghali2024gamedx} mitigate these issues, amalgamating retriever-mapped language models. Yet, this isn't a comprehensive fine-tuning solution. Our approach facilitates model fine-tuning via data retrieval, primarily leveraging vector databases for embeddings.
Contrary to conventional singular cues, recent research introduces "chaining" tasks into smaller units, enhancing performance and interpretability. Task-specific methods like chain-of-thought \cite{lievin2022can,wang2022self,tafjord2022entailer} emulate expert-like problem-solving. However, predefined prompting limits task scope. Our method employs langChain to guide Large Language Models, enabling dynamic action delegation, including internet querying or database exploration.
The crux lies in the model's knowledge-agent interplay. Retrieval augmentation \cite{lewis2020retrieval} and conversational agents synergize effectively, addressing data freshness and domain-specific knowledge gaps. The combination overcomes challenges such as agent-induced limitations. Unsophisticated retrieval augmentation 
\cite{guu2020retrieval} might overburden the model, demanding external context for every search. Yet, a balanced approach is warranted, as not all queries require external knowledge access.
On the other hand, when it comes to Text-to-SQL tasks, the integration of Large Language Models (LLMs) with database systems presents significant challenges, particularly in schema matching and translating natural language queries into SQL commands.
\cite{parameswaran2023revisiting} introduce the concept of declarative prompt engineering, drawing parallels between the operations of LLMs and human crowd workers to simplify interactions and reduce the complexity of schema matching. By treating LLMs as crowd workers, this approach leverages more structured and intuitive prompt engineering techniques, aligning natural language queries with database schemas more effectively. This methodology humanizes LLM interactions, making them more accessible and reducing the daunting task of schema matching due to the inherent variability and ambiguity in natural language.
Building on these ideas, \cite{Liu2023Divide} propose the Divide-and-Prompt paradigm, which uses the Chain-of-Thought (CoT) prompting technique to break down complex problems into smaller, manageable parts. This incremental approach allows LLMs to better understand and align with database schemas. By sequentially addressing each component of a complex query, this method improves LLMs' ability to navigate and comprehend the intricacies of database structures, thereby enhancing the accuracy and efficiency of Text-to-SQL conversions.
Data pre-processing is another critical aspect of enhancing LLM performance for Text-to-SQL tasks. \cite{Chang2023How} explore the impact of prompt construction on LLM effectiveness. Their study demonstrates that nuanced prompt engineering strategies can significantly improve the performance of LLMs by mitigating the complexities of data pre-processing. This includes normalizing natural language inputs and aligning them with the structured requirements of SQL. Carefully crafted prompts can thus serve as powerful tools to overcome data pre-processing challenges, leading to more accurate and efficient Text-to-SQL translations.
On the scalability front, \cite{Saeed2023Querying} address the broader issue of handling database queries at scale. They introduce Galois, a system designed to leverage SQL queries to extract and utilize the vast knowledge embedded within LLMs. Galois addresses scalability concerns by enabling efficient and scalable querying of databases, even when processing large volumes of queries or managing extensive databases. This approach represents a significant advancement, providing a pragmatic framework for leveraging LLM capabilities sustainably and overcoming major hurdles in Text-to-SQL tasks.
\cite{Gao2023Text-to-SQL} contribute significantly by addressing the persistent challenge of inadequate training data and models for Text-to-SQL tasks. Their detailed examination of prompt engineering techniques highlights the importance of designing prompts that enhance LLM training processes. By comparing different strategies, they reveal how nuanced adjustments in prompt design can improve LLMs' ability to learn from available datasets and accurately translate natural language queries into SQL commands. This study underscores the need for sophisticated training methodologies that leverage high-quality, diverse datasets and advance model capabilities.

The existing literature, as evidenced by the aforementioned, on the information retrieval part, the focus is on domain-specific pretrained models and their performance in specific tasks in specific domains after being trained. However, a distinct literature gap is apparent in relation to the inefficiency of storing and retrieving natural language data, particularly when conducting high-speed search operations over large datasets. The need for efficient storage and retrieval methods for natural language data is underscored, as current approaches prove impractical. This research aims to address this specific literature gap by proposing solutions for efficient storage, retrieval, and adaptable embeddings whenever you upload new material to the LLM based chatbot that will be implemented. 
On the other hand, for the efficient database querying, there remains a notable gap in efficiently processing vast amounts of data and tables without necessitating the retraining or fine-tuning of models. This is particularly costly and technically challenging. To address this, the proposed framework leverages in-context learning capabilities inherent in LLMs. This approach aims to circumvent the limitations imposed by token window constraints, which restrict the amount of information processed in a single query. By intelligently and selectively retrieving only the necessary data for query processing, this method enhances efficiency and scalability.

\section{Methodology}
The baseline model (GTR), highlighted in Figure \ref{fig1}, focuses on processing and embedding document chunks for retrieval based on semantic similarity with user queries. Recognizing the distinct characteristics and challenges of tabular data, we developed a variation (GTR-T), shown in Figure \ref{texttosql}, optimized for querying structured databases. GTR-T enhances efficiency by selectively retrieving relevant tables, thus overcoming limitations related to LLMs' context windows and enhancing overall performance in data analysis tasks across diverse domains. The methodology employed in this study integrates advanced techniques to address the challenges of information retrieval from both unstructured textual data and structured tabular data.
\subsection{Generative Text Retrieval (GTR)}
Figure \ref{fig1} shows the methodology followed in this paper starting from chunking the document to embeddings creation, loading embeddings to a vector store, then comparing these with the user query embeddings, all the way to the extraction of the most relevant chunk to be fed alongside the user query to the LLM for answer inference. The pseudocode is shown in Algorithm \ref{alg:text_embeddings} in the Appendix.

\subsubsection{Text Extraction and Embeddings Creation}
Embeddings capture semantic relationships between words. For instance, in a good embedding space, words like "man" and "women" would be closer to each other than to words like "bottle" or "paper," reflecting their semantic similarity. Similarly, embeddings can capture other relationships like gender, verb tense, and more.
In the initial phase of our methodology, we systematically process research documents to distill meaningful information. The document \(P\) undergoes a controlled chunking operation, resulting in a set of chunks \(C\). Specifically, we denote 
\newline
\begin{equation}
    C = \{c_i\}_{i=1}^{n}
\end{equation}
\newline
where each chunk \(c_i\) is generated through the operation \(\text{Chunk}_i(P)\).
Subsequently, employing embedding creation algorithms such as Word2Vec, each chunk is transformed into semantic embeddings denoted by
\newline
\begin{equation}
    E(c_i) = \text{Vectorize}(c_i)
\end{equation}
\newline
This meticulous process captures the nuanced semantic relationships within the content, yielding a set of embeddings \(E(c_i)\) that collectively encapsulate the underlying meaning of each chunk.

\subsubsection{Embeddings Storing and Retrieval from Vector Database}
Through a careful process of encoding and representation, the textual content extracted from research documents, they will be transformed into high-dimensional vectors. These vectors will encapsulate the underlying semantic meaning, contributing to the chatbot's ability to comprehend user queries and retrieve relevant information.
Our methodology centers on the storage and retrieval of embeddings within a dedicated vector database. The vector database \(V\) is systematically constructed as 
\newline
\begin{equation}
    V = \{E(c_i)\}_{i=1}^{n}
\end{equation}
\newline
where each \(E(c_i)\) originates from the text extraction phase.
Upon receiving a user query \(Q\), the embeddings of the query (\(E(Q) = \text{Embedding}(Q)\)) are methodically compared with each stored embedding. Specifically, the cosine similarity measure is employed, resulting in a set of similarities denoted as 
\newline
\begin{equation}
    \text{Similarities} = \{\text{Sim}(E(Q), E(c_i))\}_{i=1}^{n}
\end{equation}
\newline
The index with the highest similarity, identified as 
\newline
\begin{equation}
    \text{Index} = \arg\max_i \text{Sim}(E(Q), E(c_i))
\end{equation}
\newline
corresponds to the most contextually relevant chunk in the vector database. Let \(L\) represent the large language model. 
\newline
\begin{equation}
\newline
    \text{GTR Output} = E(Q) + E(c_{\text{i}})
\end{equation}
\newline
This selected chunk, combined with the user query, serves as the input for subsequent interactions with the large language models, ensuring an optimized and contextually accurate retrieval process.

\begin{figure}[H]
	\centering
	\includegraphics[width=0.9\textwidth]{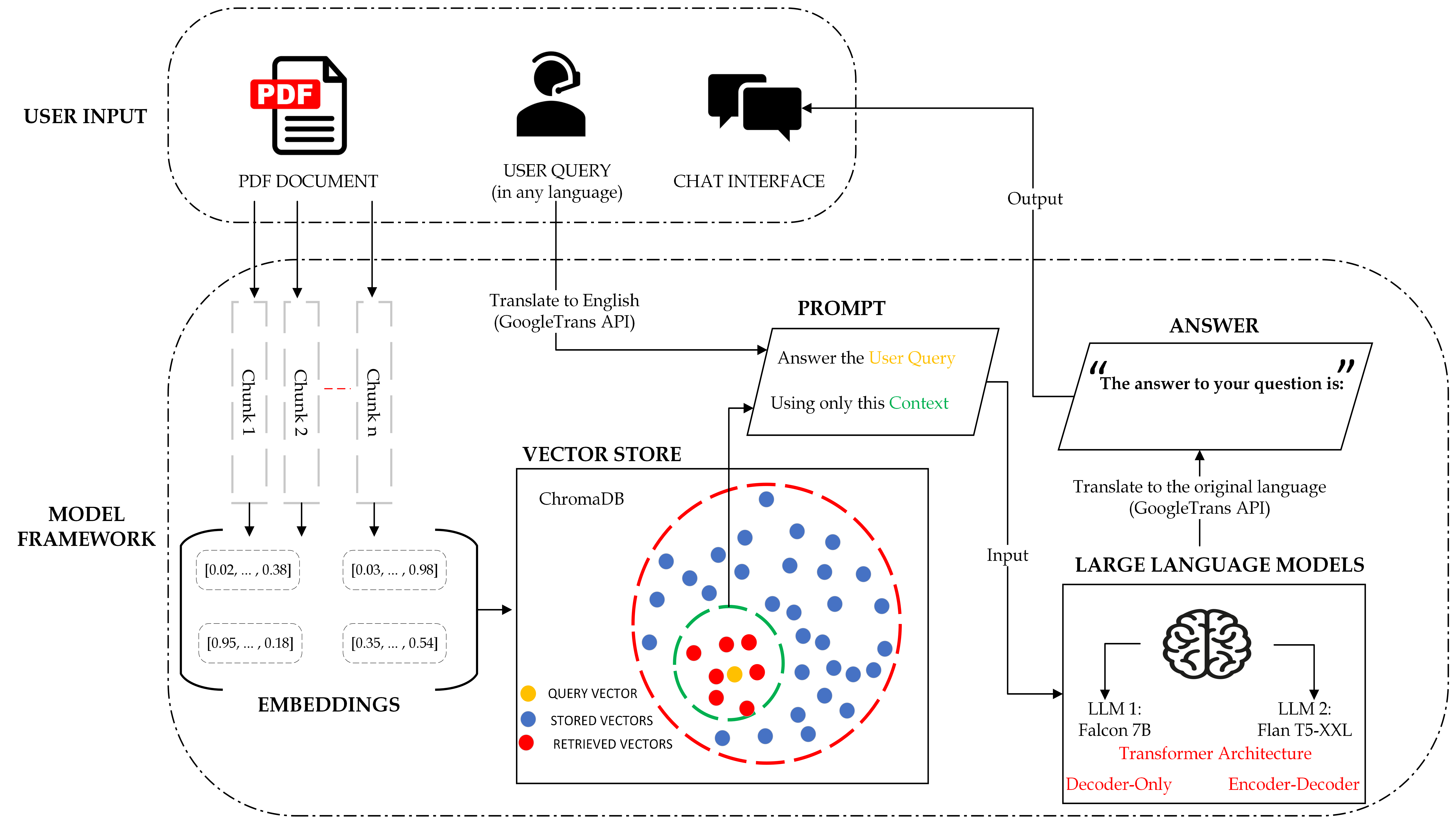}
	\caption{Methodology overview for GTR showing our embedding-based retrieval system using document chunking, vector database storage, and LLM inference for retrieval of relevant textual data and formulation of answer}\label{fig1}
\end{figure}

\subsection{Generative Tabular Text Retrieval (GTR-T)}
\begin{figure}[H]
	\centering
	\includegraphics[width=0.9\textwidth]{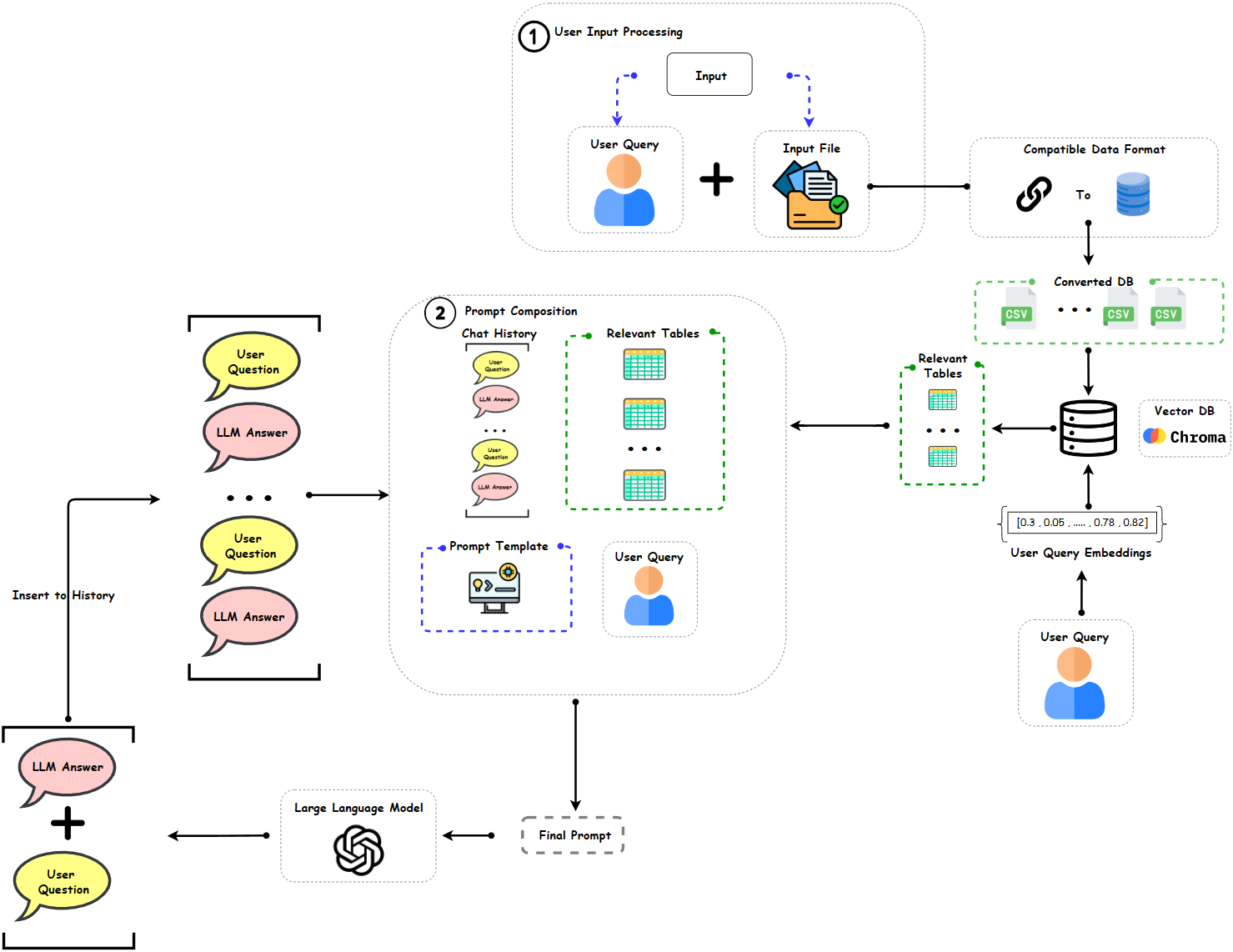}
	\caption{Methodology overview for GTR-T highlighting our approach for structured data retrieval, utilizing table embeddings and LLM generated SQL query to efficiently query large databases}\label{texttosql}
\end{figure}

Based on the idea described in figure \ref{fig1}, a variation of this algorithm can be used to query large databases, since LLMs struggle with limited context-window, retrieving only relevant tables to execute the query on will cut off both token usage and execution time.It aims to enhance the efficiency of data analysis across various tables within a database using Large Language Models (LLMs). The process initiates with retrieving database tables and their metadata, converting them into CSV format for ease of processing, and storing them in a vector database for efficient retrieval. When a user query is received, its embeddings are computed to capture semantic meaning. These embeddings are then compared with those of stored tables to identify relevant ones. The selected tables, along with the query, are combined into a coherent prompt for the LLM. The LLM processes this prompt to generate an SQL query, which is executed on the relevant tables to fetch the desired information. This methodology streamlines the process of querying structured/tabular data within a database by leveraging the capabilities of LLMs for natural language understanding and SQL generation. Figure \ref{texttosql} shows the GTR-T architecture. The algorithm for this methodology \ref{alg:data_analysis} is highlighted in the Appendix.

\subsubsection{Query Processing and Relevant Tables Mapping}
Data Retrieval and Conversion: Let \( T_{DB} \) represent the set of tables retrieved from the database. To transform these tables into a CSV format suitable for further processing, each table \( T_i \) is converted into CSV format denoted as \( T_{CSV}^i \). Mathematically, this can be expressed as:
\begin{equation}
T_{CSV}^i = \text{ConvertToCSV}(T_i) \quad \text{for} \quad T_i \in T_{DB}
\end{equation}
Vector Database Storage: Following the conversion, the CSV-formatted tables \( T_{CSV}^i \) are vectorized stored in a vector database, \( VDB \). This storage operation can be represented as:
\begin{equation}
 E(T_{CSV}^i) = \text{Vectorize}(T_i) \quad \text{for} \quad T_i \in T_{CSV}^i
\end{equation}
User Query Processing: Upon receiving a user query \( Q \), embeddings are computed to capture its semantic meaning and context. Let \( E_Q \) denote the embeddings for the user query. This process can be represented mathematically as:
\begin{equation}
E_Q = \text{Vectorize}(Q)
\end{equation}
Similarity Computation: Computed embeddings for the user query (\( E_Q \)) are then compared with the embeddings of tables stored in the vector database (\( E_{T_{CSV}^i} \)) to identify the most relevant tables. Utilizing cosine similarity for comparison:
\begin{equation}
\text{Similarity}(E_Q, E_{T_{CSV}^i}) = \frac{{E_Q \cdot E_{T_{CSV}^i}}}{{\|E_Q\| \|E_{T_{CSV}^i}\|}}
\end{equation}
This computation yields a similarity score between the user query and each table, facilitating the identification of relevant tables based on their similarity to the query.
\subsubsection{Prompt Composition and Injection to the LLM }
When the most relevant tables, along with the user query, are selected and chained together to form a coherent prompt. This prompt encapsulates the necessary information for the LLM to process the query effectively. This can be expressed as:
\begin{equation}
\text{Prompt} = \text{ChainTablesAndQuery}(T_{\text{relevant}}, Q)
\end{equation}
LLM Inference: The prompt, comprising the user query and relevant tables, is passed to the LLM for inference. LLMs, known for their proficiency in text-to-SQL tasks, generate the SQL query corresponding to the user's intent. This process can be represented as:
\begin{equation}
\text{SQLQuery} = \text{LLMInference}(\text{Prompt})
\end{equation}
Query Execution and Result Retrieval: The generated SQL query is executed on the database, fetching the desired information. The final answer or result is then provided to the user, completing the data analysis process. This can be expressed as:
\begin{equation}
\text{Result} = \text{ExecuteSQLQuery}(\text{SQLQuery})
\end{equation}

\section{Experimental Setup}
In addition to MSMARCO \cite{bajaj2018ms} evaluation dataset, a set of 15 carefully crafted questions was generated to evaluate the Large Language Models (LLMs)' comprehension and responsiveness. Among these, 10 questions were thoughtfully aligned with the thematic content of the knowledge documents, while the remaining 5 questions were intentionally designed to be unrelated. To simulate real-world interactions, each question was meticulously formatted as a natural inquiry directed towards the chatbot. The experimental execution involved the integration of the chatbot system with LangChain and a vector database. Each question was sequentially processed, and responses were generated using the two LLMs. The evaluation process was underpinned by a set of robust metrics, including ROUGE and Semantic Answer Similarity (SAS) accuracy, response token length, speed of response, and truthfulness. For accuracy assessment, responses were scrutinized for relevance to the original question, and accuracy was measured using both ROUGE and SAS.
ROUGE is a set of metrics, including ROUGE-N (measuring n-gram overlap) and ROUGE-L (measuring longest common subsequence), used in text summarization evaluation to quantitatively assess the precision of the summary by comparing it to reference documents. It provides a technical basis for measuring the quality and informativeness of generated summaries~\cite{lin2004rouge}.
The SAS metric assesses the semantic similarity between model-generated answers and ground-truth annotations, addressing the limitation of lexical-based metrics. It is important in text summarization evaluation as it recognizes semantic similarity, enhancing model assessment over traditional metrics that focus on lexical overlap and may overlook valid but lexically distinct answers~\cite{risch2021semantic}.
The token length of documents was measured to provide insights into their conciseness or verbosity. Additionally, the time taken by the chatbot to generate responses was recorded to gauge response speed. A dichotomous truthfulness rating, assigning 1 for relevant and 0 for irrelevant responses, further contributed to the comprehensive evaluation.

\begin{table}[h]
 \caption{Models Basic Comparison}
    \label{tab:models-comparison}
    \centering
    \small 
    \vspace{-\abovecaptionskip}
    \begin{tabular}{lccc}
        \hline
        \textbf{Model} & \textbf{Model Size} & \textbf{Base Architecture} & \textbf{Max Sequence Length (tokens)} \\
        \hline
        Falcon 7B & 7B parameters & Decoder-Only & 2048 \\
        Flan T5-XXL & 11B parameters & Encoder-Decoder & 2048 \\
        \hline
    \end{tabular}
\end{table}
Table \ref{tab:models-comparison} provides a concise summary of the language models employed in the experimental analysis detailed in this paper. The models under consideration are Falcon 7B and Flan T5-XXL. Falcon 7B utilizes a Decoder-Only architecture optimized for tasks involving text generation and completion. These models exhibit proficiency in processing sequences up to a length of 2048 tokens. On the other hand, Flan T5-XXL, an expansive model with 11 billion parameters, adopts an Encoder-Decoder framework, enabling comprehensive handling of input sequences and subsequent generation of corresponding outputs, all within a maximum sequence length of 2048 tokens.

We assessed the performance of our refined model, GTR-T, using the Spider dataset, which comprises a total of 10,181 questions and 5,693 distinct complex SQL queries. These queries span across 200 databases, covering 138 different domains, each containing multiple tables.
The dataset follows a standardized protocol, dividing the examples into training, development, and test sets. Specifically, there are 8,659 training examples spread across 146 databases, 1,034 development examples across 20 databases, and 2,147 test examples across 34 databases. Importantly, the databases used in each of these sets do not overlap.
To gauge the complexity of SQL queries, they are categorized into four difficulty levels. This categorization is based on factors such as the number of SQL keywords employed, the presence of nested subqueries, and the utilization of column selections and aggregations.

We used the official evaluation metrics provided by the Spider Dataset: Exact-Set-Match Accuracy (EM) and Execution Accuracy (EX).
\begin{itemize}
    \item \textbf{Exact-Set-Match Accuracy (EM):} considers each clause within a SQL query as a set and compares the prediction for each clause against the corresponding clause in the reference query. To be deemed correct, a predicted SQL query must match the ground truth query in all its components, disregarding the actual values involved. This metric emphasizes the structural accuracy of the generated queries.
    \item \textbf{Execution Accuracy (EX):} evaluates the predicted SQL query's execution output against that of the ground truth query on specific database instances. This metric offers a more accurate evaluation of the model's performance, accounting for scenarios where there might be multiple valid SQL queries. However, this metric is still limiting in terms of semantic comparison.
\end{itemize}

\section{Results}
\subsection{Evaluation of GTR}
Table \ref{tab:truthfulness-comparison} illustrates the truthfulness assessment of the 2 models, Falcon 7B and Flan T5-XXL, It highlights the frequency of relevant (1) and irrelevant (0) responses for each model, revealing their efficacy in providing accurate information within the context of the experiment.
\begin{table}[H]
\caption{Truthfulness Comparison}
    \label{tab:truthfulness-comparison}
    \centering
    \small 
    \vspace{-\abovecaptionskip}
    \begin{tabular}{lcc}
        \hline
        \textbf{Model} & \textbf{Truthful Count (\%)} & \textbf{Untruthful Count (\%)} \\
        \hline
        Falcon 7B & 73.00\% & 27.00\% \\
        Flan T5-XXL & 87.00\% & 13.00\% \\
        \hline
    \end{tabular}  
\end{table}
It is evident that Flan T5-XXL outperforms Falcon 7B in terms of producing truthful content, with an accuracy rate of 87\%, while Falcon 7B achieves 73\%. Conversely, Falcon 7B generates untruthful responses 27\% of the time, compared to 13\% for Flan T5-XXL. These results underscore Flan T5-XXL's higher overall accuracy in generating truthful content, making it a more reliable choice for applications where factual accuracy is crucial.

\begin{table}[H]
\caption{Key Metrics Comparison}
    \label{tab:key-metrics-comparison}
    \centering
    \small  
    \vspace{-\abovecaptionskip}
    \begin{tabular}{lccc}
        \hline
        \textbf{Model} & \textbf{Average Accuracy (\%)} & \textbf{Documents Token Length} & \textbf{Average Response Time (s)} \\
        \hline
        Falcon 7B & 80.50\% & 3890 & 5.100 \\
        Flan T5-XXL & 92.50\% & 3890 & 2.000 \\
        \hline
    \end{tabular}
    
\end{table}
Table \ref{tab:key-metrics-comparison} presents key metrics for each model's performance on specific questions. It showcases average accuracy ratings (80.50\%, 92.50\%), token lengths of 3890, and response times (5.1s, 2.0s), offering insights into their responsiveness and effectiveness in real-world inquiries.
In terms of average accuracy, Flan T5-XXL outperforms Falcon 7B, achieving an accuracy of 92.5\%, while Falcon 7B scores 80.5\%. Both models were compared on questions from a source document with the same token length of 3890. When considering response time, Flan T5-XXL demonstrates a quicker average response time of 2.0 seconds, in contrast to Falcon 7B, which takes 5.1 seconds on average to generate responses. These findings highlight Flan T5-XXL's superiority in terms of accuracy and response time efficiency, which can be advantageous for applications requiring both accuracy and speed.
\begin{table}[H]
\caption{Linguistic Metrics Comparison (ROUGE and SAS)}
    \label{tab:linguistic-metrics-comparison}
    \centering
    \small 
    \vspace{-\abovecaptionskip}
    \begin{tabular}{lcccc}
        \hline
        \textbf{Model} & \textbf{SAS Score} & \textbf{Rouge-1 Precision} & \textbf{Rouge-2 Precision} & \textbf{Rouge L-Precision} \\
        \hline
        Falcon 7B & 0.953 & 0.711 & 0.551 & 0.656 \\
        Flan T5-XXL & 0.936 & 0.929 & 0.871 & 0.929 \\
        \hline
    \end{tabular}
\end{table}
The results indicate that Falcon 7B model achieved a relatively high SAS score, suggesting a strong ability to capture semantic similarity. However, its ROUGE scores were lower, implying potential limitations in lexical precision. In contrast, Flan T5-XXL model demonstrated excellent ROUGE scores, especially for Rouge-2, signifying strong lexical precision, but its SAS score was slightly lower. These findings emphasize the trade-off between semantic and lexical performance in LLMs, highlighting the importance of selecting the right model based on specific summarization needs and objectives.

\begin{figure}[H]
  \centering
  \begin{subfigure}[b]{0.49\textwidth}
    \includegraphics[width=\textwidth]{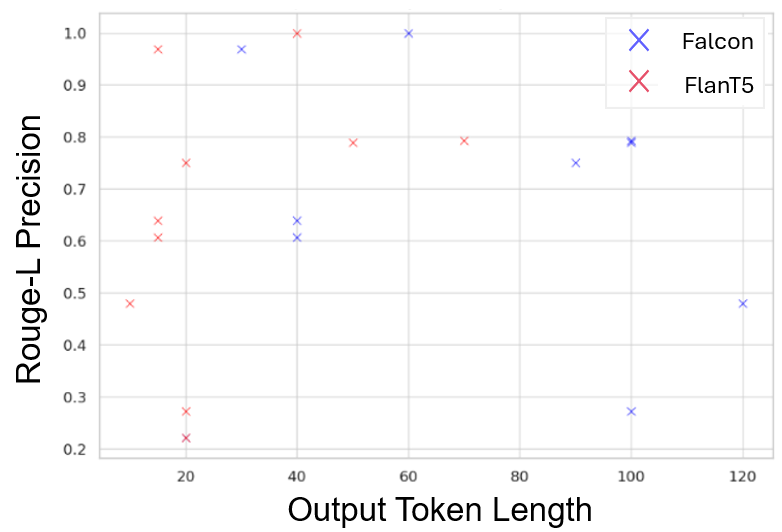}
    \caption{}
    \label{fig:precisiontokenlength}
  \end{subfigure}
  \hfill
  \begin{subfigure}[b]{0.49\textwidth}
    \includegraphics[width=\textwidth]{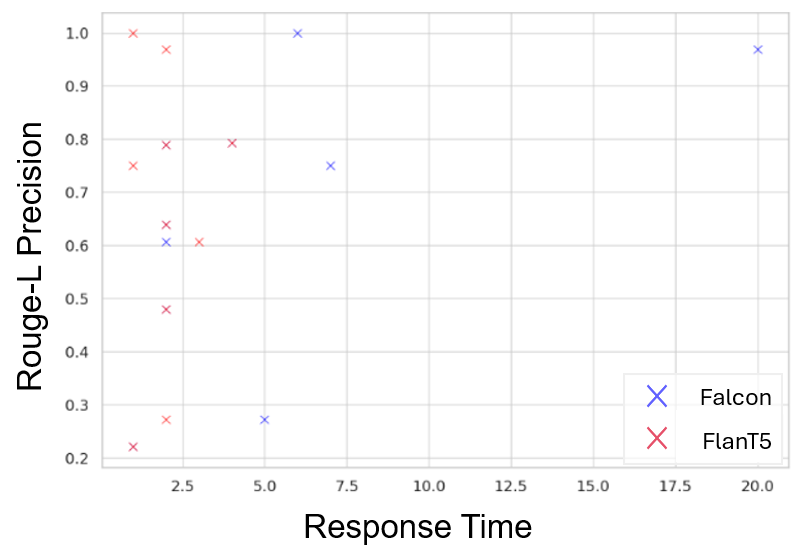}
    \caption{}
    \label{fig:precisionresponsetime}
  \end{subfigure}
  \caption{ GTR Reponse and Output Length}
\end{figure}

Investigating detailed results beyond simple averages is crucial for a comprehensive understanding of model performance for several reasons. First, averages can mask underlying variability and extreme values, potentially overlooking instances where the model performs exceptionally well or poorly. For that, we are studying in graphs \ref{fig:precisiontokenlength} and \ref{fig:precisionresponsetime} if there are any underlying relationships between the precision of the response(quality of response) and key metrics such as output token length and average response time.

Figure \ref{fig:precisiontokenlength} helps to understand whether the length of the output affects its quality. The distribution of points for both models shows no clear trend, suggesting that the quality of the generated text, as measured by Rouge-L Precision, is relatively stable across different output lengths. This stability is beneficial as it indicates that both models can produce high-quality text whether the output is short or long. However, the presence of scattered points across the plot highlights variability in text quality at various lengths. Analyzing these outliers could provide insights into specific cases where the models excel or struggle, potentially guiding improvements in model performance or identifying optimal use cases for each model based on the desired output length.

Figure \ref{fig:precisionresponsetime} indicates that the quality of the generated text is not strongly correlated with the models' response times. The scattered points across various response times without a clear pattern suggest that both models can maintain the quality of the text they produce regardless of how quickly they respond. This is desirable, as it means that efficiency (in terms of response time) does not compromise text quality.

However, observing some points that significantly deviate from the main cluster in both Figures \ref{fig:precisionresponsetime} and \ref{fig:precisiontokenlength} might highlight specific instances where either model performs unusually well or poorly, which could be worth investigating further.
For this purpose, two extra embeddings models were used to evaluate the output of the two models, to see how the questions and answers are distributed in the embeddings space, t-SNE space reduction technique \cite{van2008visualizing} was used to reduce the embeddings space dimensionality to two dimensions for an easier interpretability. 
\begin{figure}[H]
  \centering
  \begin{subfigure}[b]{0.49\textwidth}
    \includegraphics[width=\textwidth]{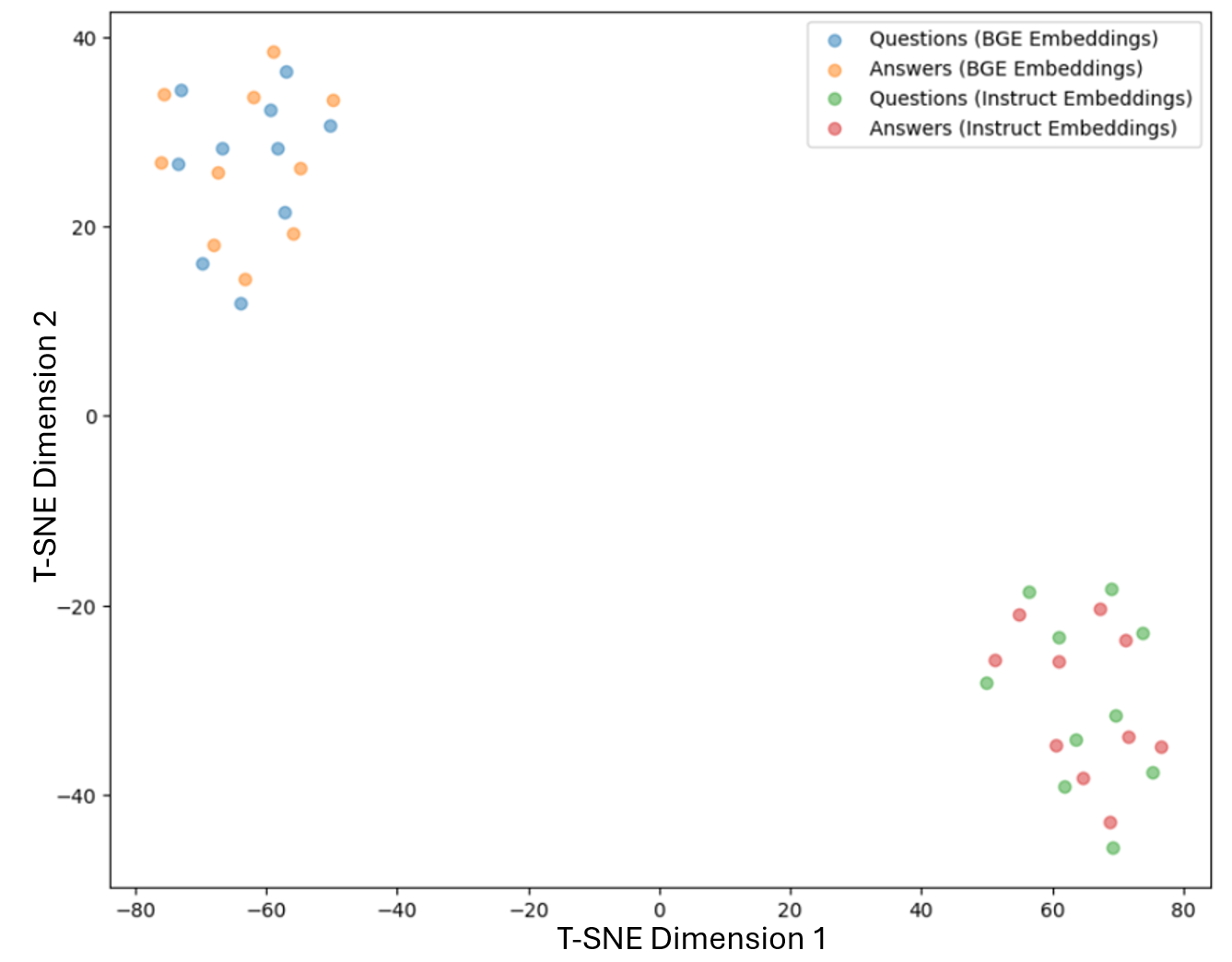}
    \caption{}
    \label{fig:falcontsne}
  \end{subfigure}
  \hfill
  \begin{subfigure}[b]{0.49\textwidth}
    \includegraphics[width=\textwidth]{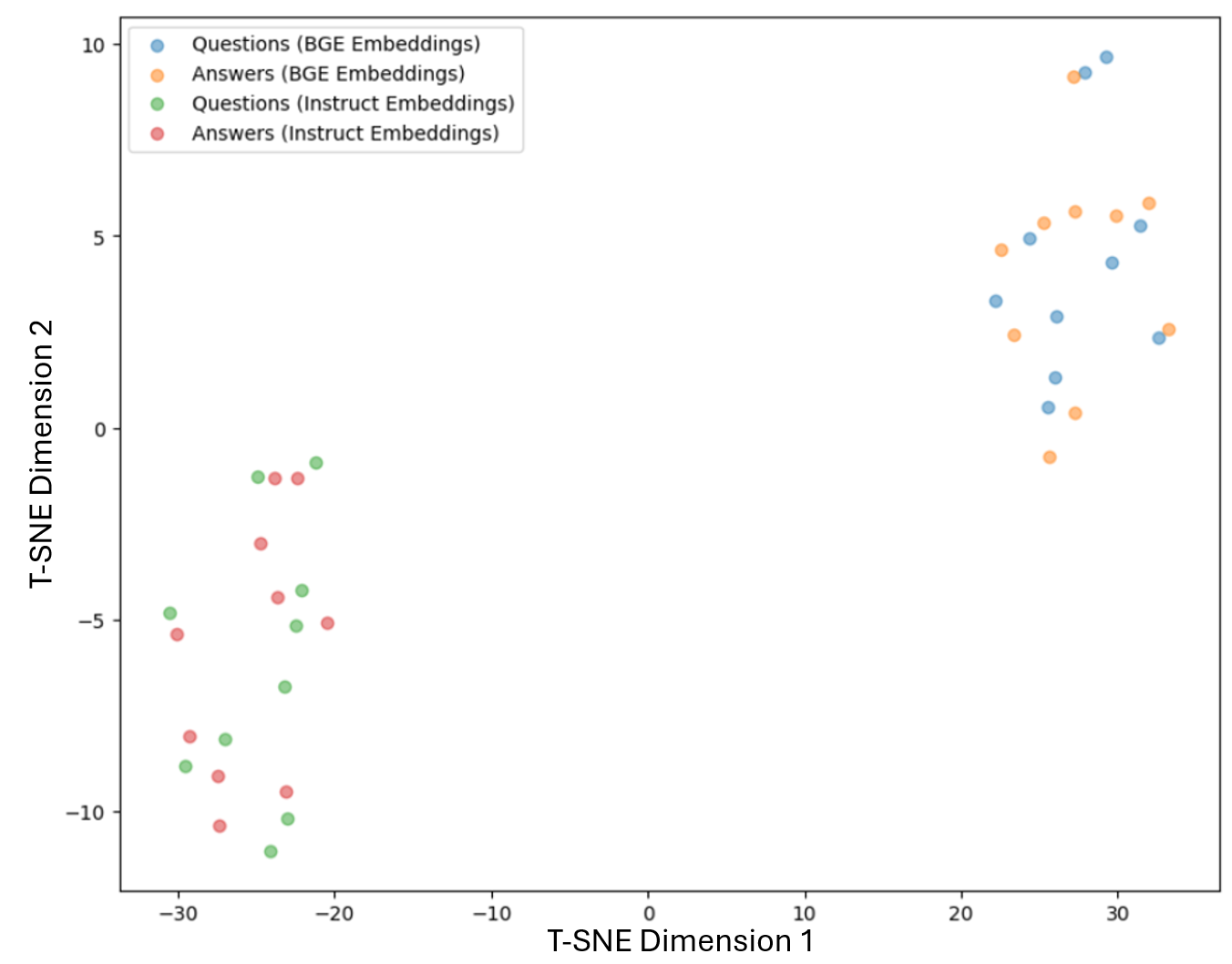}
    \caption{}
    \label{fig:flantsne}
  \end{subfigure}
  \caption{GTR Semantic Evaluation}
\end{figure}

 The t-SNE plots \ref{fig:falcontsne} and \ref{fig:flantsne} shows a clear distinction between the clusters of questions and answers, indicating that the embeddings capture distinct features of questions and answers, with both embeddings models for both questions and answers are closer together, suggesting a tighter semantic relationship or less variability within the embedding space for the Falcon model. The tighter clusters, particularly for the BGE Embeddings, suggest that the Flan T5 model generates embeddings with less variability between questions and answers, possibly pointing to a more consistent generation approach. The consistency in representation across different embedding types is a strong indicator that the methodology proposed is not overly sensitive to the LLM at the backend and, thus, can be reliably used with any LLM at the same level of tokens.

\begin{table}[htbp]
\centering
\caption{Rouge-L scores of different models on the MSMARCO dataset {\textit{\tiny{a.}}}\cite{seo2018bidirectional}, {\textit{\tiny{b.}}}\cite{Yan_2019}, {\textit{\tiny{c.}}}(Unpublished), {\textit{\tiny{d.}}}(Unpublished), {\textit{\tiny{e.}}}(Unpublished), {\textit{\tiny{f.}}}\cite{wang2018multipassage}, {\textit{\tiny{g.}}}Ours. }
\label{tab:rouge_scores}
\begin{tabular}{@{}ll@{}}
\toprule
\textbf{\textbf{Model}}              & \textbf{Rouge-L} \\ \midrule
\textbf{\textit{\tiny{a}}} BiDAF                       & 0.2396                  \\
\textbf{\textit{\tiny{b}}} Deep Cascade QA             & 0.5201                  \\
\textbf{\textit{\tiny{c}}} S-Net+CES2S                 & 0.4496                  \\
\textbf{\textit{\tiny{d}}} BERT+Multi-PGNet            & 0.4814                  \\
\textbf{\textit{\tiny{e}}} VNET                        & 0.5163                  \\
\textbf{\textit{\tiny{f}}} Masque (Q\&A; ensemble)     & 0.5220                  \\
 \textbf{\textit{\tiny{g}}} \textbf{GTR}                    & \textbf{0.9870}                  \\ \bottomrule
\end{tabular}
\end{table}
In the assessment of question-answering models using the MSMARCO evaluation dataset, a range of models underwent evaluation through Rouge-L scores. As shown in the results highlighted in Table \ref{tab:rouge_scores}, while certain models showed moderate to satisfactory performance, such as BiDAF registering a score of 0.2396 and BERT+Multi-PGNet achieving 0.4814, others demonstrated significant advancements, like Deep Cascade QA with a score of 0.5201 and Masque (Q\&A; ensemble) hitting 0.5220. Nonetheless, standing out among them all, my model showcased exceptional ability with a Rouge-L score of 0.9870, underscoring the effectiveness of its LLM-based approach in comprehensively grasping and accurately addressing questions based on diverse and complex passages.
\subsection{Evaluation of GTR-T}
The results presented in Table \ref{tab:t2sql} reveal the performance of various Text2SQL models, in comparison with GTR-T. GTR-T exhibits competitive performance in both Execution Accuracy (EX) and Exact-Set-Match Accuracy (EM). With an EX Accuracy of 0.820, GTR-T successfully generates SQL queries that produce the correct execution output approximately 82\% of the time when compared to ground truth queries on specific database instances. Moreover, its EM Accuracy of 0.597 indicates that nearly 60\% of the predicted SQL queries precisely match the reference queries in terms of structural components, regardless of specific values. Comparatively, DIN-SQL(4), powered by GPT-4, achieves slightly lower EM Accuracy but with a similar EX Accuracy to GTR-T. C3, fueled by GPT-3, demonstrates inferior performance in both EM and EX Accuracy compared to GTR-T. While DAIL-SQL(4) performs slightly better than GTR-T in terms of both EM and EX Accuracy, the results collectively suggest that GTR-T, powered by the Mistral model, competes effectively with existing Text2SQL models in accurately translating natural language questions into SQL queries for database retrieval tasks.
\begin{table}[H]
\centering
\caption{EX and EM Accuracy for Text2SQL Models {\textit{\tiny{a.}}}\cite{Gao2023Text-to-SQL}, {\textit{\tiny{b.}}}\cite{pourreza2023dinsql}, {\textit{\tiny{c.}}}\cite{dong2023c3}, {{\textit{\tiny{d.}}} Ours}}
\label{tab:t2sql}
\begin{tabular}{lccc}
\hline
\textbf{\textbf{Model}} & \textbf{EX Accuracy} & \textbf{EM Accuracy} \\ \hline
\textbf{\textit{\tiny{a}}} \textbf{DAIL-SQL(4)}             & \textbf{0.836}               & \textbf{0.687}                     \\
\addlinespace
\textbf{\textit{\tiny{b}}} DIN-SQL(4)              & 0.828               & 0.601                     \\
\addlinespace
\textbf{\textit{\tiny{c}}} C3                      & 0.818               & 0.431                     \\
\addlinespace
\textbf{\textit{\tiny{d}}} GTR-T               & 0.820            & 0.597                    \\ \hline
\end{tabular}
\end{table}
\par
\begin{figure}[H]
    \centering
    \begin{subfigure}[b]{0.3\textwidth}
        \includegraphics[width=\textwidth]{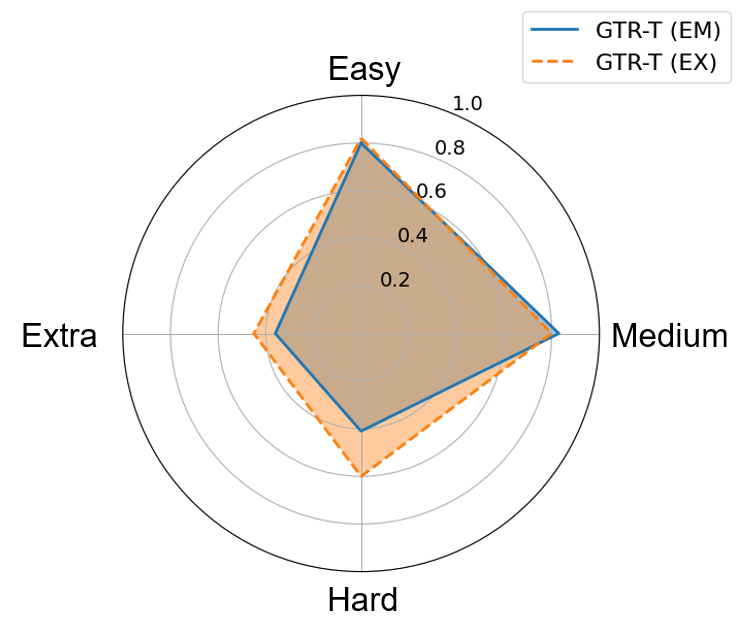}
        \caption{}
        \label{fig:spider}
    \end{subfigure}
    \hfill
    \begin{subfigure}[b]{0.3\textwidth}
        \includegraphics[width=\textwidth]{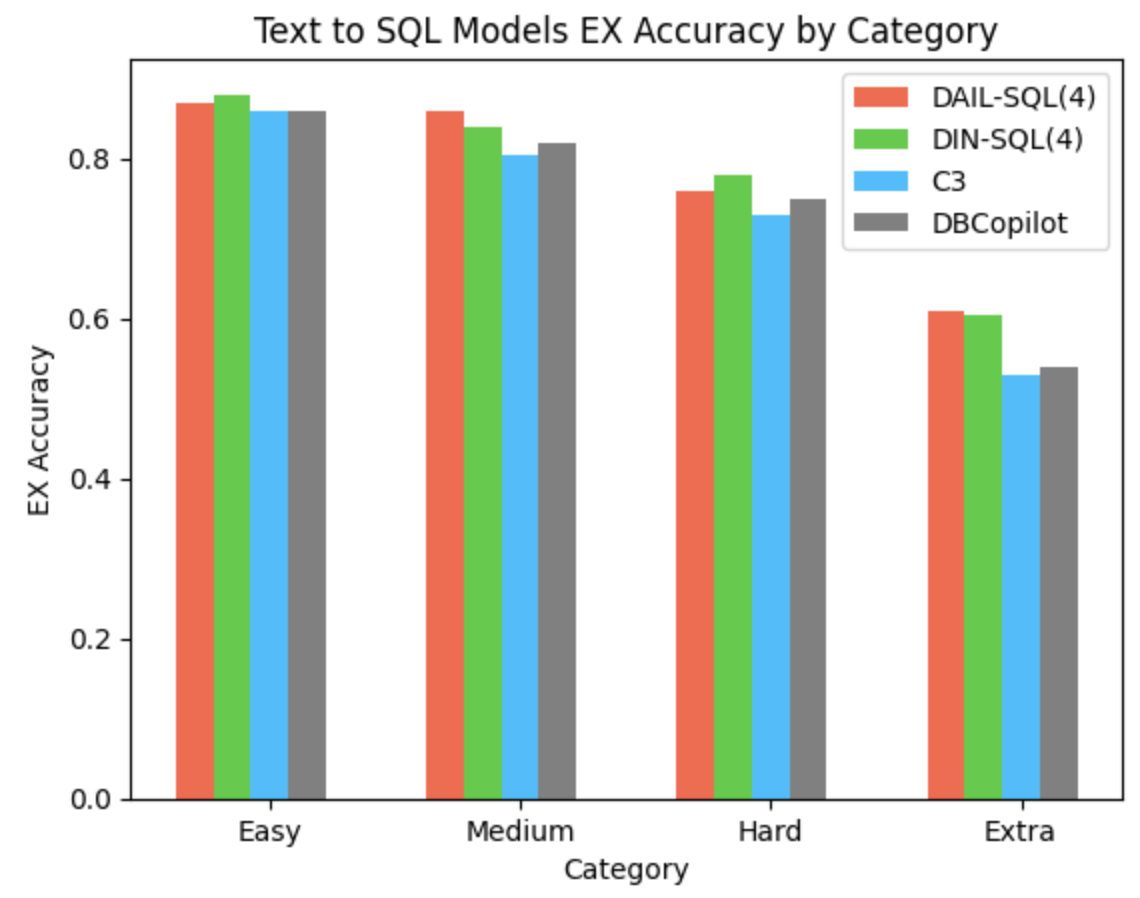}
        \caption{}
        \label{fig:exacc}
    \end{subfigure}
    \hfill
    \begin{subfigure}[b]{0.3\textwidth}
        \includegraphics[width=\textwidth]{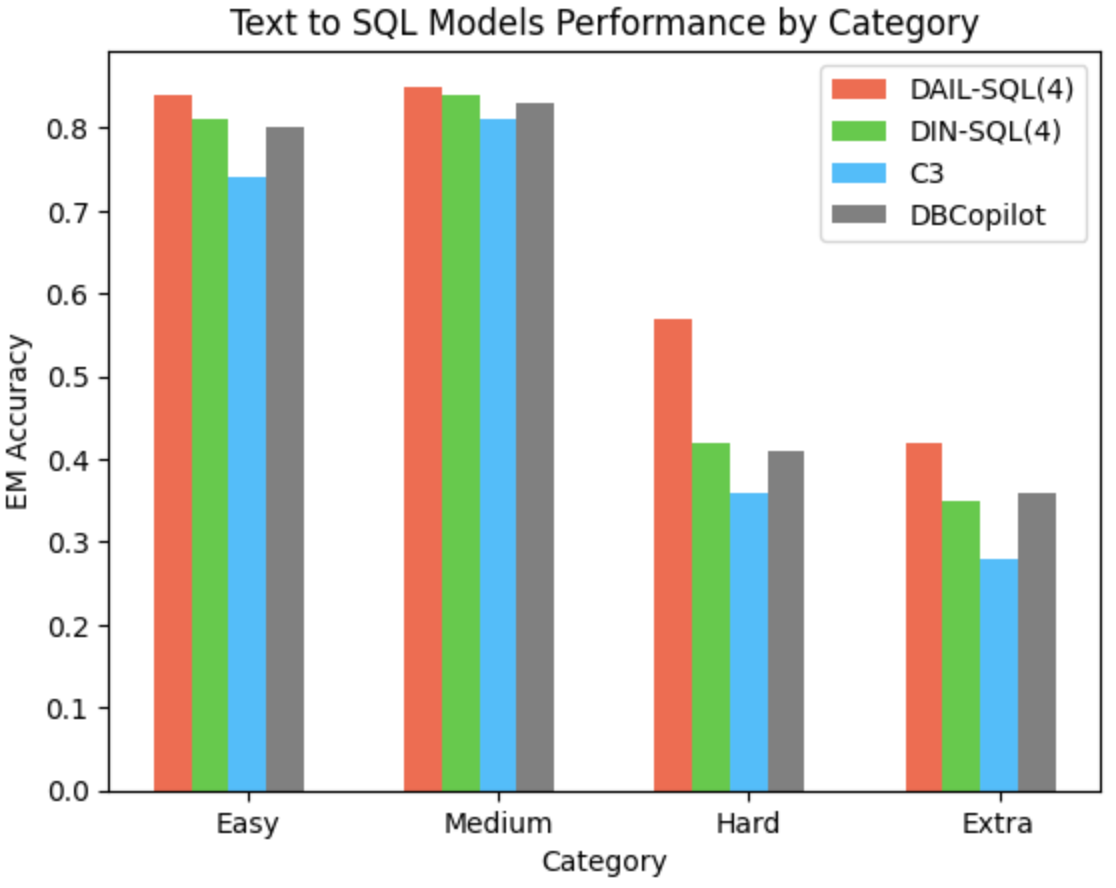}
        \caption{}
        \label{fig:emacc}
    \end{subfigure}
    \caption{Performance Metrics of GTR-T Model}
    \label{fig:combined}
\end{figure}

 Analyzing both the Exact-Set-Match Accuracy (EM) and Execution Accuracy (EX) across difficulty categories as shown in Figures \ref{fig:spider}, \ref{fig:emacc} and \ref{fig:exacc}  provides a deeper evaluation the performance of Text2SQL models. DAIL-SQL(4) consistently performs well in EM Accuracy across all difficulty levels, especially in easier and medium categories. DIN-SQL(4) and GTR-T also demonstrate competitive EM Accuracy scores, indicating their proficiency across various difficulty levels. However, C3 consistently lags behind, particularly in harder and very hard categories. In terms of Execution Accuracy, DIN-SQL(4) maintains a lead across all difficulty levels, closely followed by DAIL-SQL(4) and GTR-T, all achieving high EX Accuracy. Conversely, C3 shows a noticeable decline in Execution Accuracy, especially in harder categories, suggesting challenges in generating accurate SQL queries. While DAIL-SQL(4) and DIN-SQL(4) show robust performance across both metrics, GTR-T competes well, particularly in Execution Accuracy. However, C3 faces challenges in maintaining accuracy across difficulty levels.

\section{Conclusions}
In conclusion, this work introduces a novel approach to tackle the challenges posed by the rapid increase in data within the realm of Generative AI. By proposing an In-Context Learning-centered methodology with large language models (LLMs). This study pioneers a new framework in human-text interaction through the integration of LLMs and In-Context Learning, aiming to enhance processes such as information retrieval for both structured and unstructured data through GTR and GTR-T. By addressing limitations inherent in LLMs, such as difficulties with domain-specific data and restricted context windows, the proposed methodologies enable users to leverage advanced Generative AI capabilities confidently, without requiring deep technical expertise. Additionally, the thorough evaluation of these multi-field methodologies ensures they provide cutting-edge solutions, positioning them at the leading edge of AI application. Ultimately, this innovative approach aims to democratize Generative AI, making it more accessible and impactful for everyone.
By prioritizing the accuracy and efficiency of generated responses, GTR model have demonstrated a high accuracy rate and quick response times, which can significantly improve user experience and the reliability of outputs with limited hallucinations. Additionally, the incorporation of comprehensive evaluation metrics, such as EX, EM, ROUGE and Semantic Answer Similarity, provides a well-rounded assessment of the quality of generated content.

\newpage


\newpage
\section*{Appendix}
\begin{minipage}{\linewidth}
\begin{algorithm}[H]
\caption{Generative Text Retrieval}
\label{alg:text_embeddings}
\begin{algorithmic}
\Procedure{GTR}{$P$, $Q$, $VDB$}
    \State \textbf{Input:} Research  document $P$, user query $Q$, vector database $VDB$
    \State \textbf{Output:} Relevant embeddings $E(Q)$ or $E(c_{\text{i}})$
    
    \State \textbf{Phase 1: Text Extraction and Chunking}
    \State Chunk the  document $P$ into segments: $C = \{\text{Chunk}_i(P)\}_{i=1}^{n}$
    
    \State \textbf{Phase 2: Embeddings Creation}
    \For{each chunk $c_i \in C$}
        \State Compute semantic embeddings: $E(c_i) \gets \text{Vectorize}(c_i)$
        \State Store $E(c_i)$ in vector database $VDB$
    \EndFor
    
    \State \textbf{Phase 3: User Query Processing}
    \State Compute embeddings for user query $Q$: $E(Q) \gets \text{Vectorize}(Q)$
    
    \State \textbf{Phase 4: Retrieval of Relevant Embeddings}
    \State Initialize $max\_similarity \gets 0$, $c_{\text{Index}} \gets \text{null}$
    \For{each $E(c_i) \in VDB$}
        \State Compute cosine similarity: $similarity \gets \frac{E(Q) \cdot E(c_i)}{\|E(Q)\| \|E(c_i)\|}$
        \If{$similarity > max\_similarity$}
            \State $max\_similarity \gets similarity$
            \State $c_{\text{Index}} \gets c_i$
        \EndIf
    \EndFor
    
    \State \textbf{Phase 5: Model Input Construction}
    \State GTR Output: $Output \gets E(Q) + E(c_{\text{i}})$
    
    \State \textbf{return} $Output$
    \State \textbf{Phase 6: LLM Inference}
    \State Final LLM Response: $Final Result \gets LLM(Output$)
\EndProcedure
\end{algorithmic}
\end{algorithm}
\end{minipage}

\begin{minipage}{\linewidth}
\begin{algorithm}[H]
\caption{Generative Tabular Text Retrieval }
\label{alg:data_analysis}
\begin{algorithmic}[]
\Procedure{GTR-T}{$Q$, $\mathcal{DB}$}
    \State Retrieve tables $T_{DB}$ from $\mathcal{DB}$
    \For{each table $T_i \in T_{DB}$}
        \State Convert $T_i$ into CSV format: $T_{CSV}^i \gets \text{ConvertToCSV}(T_i)$
        \State Compute and store embeddings of tables: $E(T_{CSV}^i) \gets \text{Vectorize}(T_i)$
        \State Store $T_{CSV}^i$ in vector database $VDB$
    \EndFor
    \State Compute embeddings for user query $Q$: $E_Q \gets \text{Vectorize}(Q)$
    \State Initialize $max\_similarity \gets 0$
    \For{each table $T_{CSV}^i$ in $VDB$}
        \State Compute similarity: $similarity \gets \frac{{E_Q \cdot E_{T_{CSV}^i}}}{{\|E_Q\| \|E_{T_{CSV}^i}\|}}$
        \If{$similarity > max\_similarity$}
            \State $max\_similarity \gets similarity$
            \State $T_{relevant} \gets T_{CSV}^i$
        \EndIf
    \EndFor
    \State Construct prompt: $Prompt \gets \text{ChainTablesAndQuery}(T_{relevant}, Q)$
    \State Generate SQL query: $SQL\_Query \gets \text{LLMInference}(Prompt)$
    \State Execute $SQL\_Query$ on $\mathcal{DB}$ to fetch desired information: $Result \gets \text{ExecuteSQLQuery}(SQL\_Query)$
    \State \textbf{return} $Result$
\EndProcedure
\end{algorithmic}
\end{algorithm}
\end{minipage}

\end{document}